\documentclass[10pt]{iopart}

\usepackage{iopams}
\usepackage{cite}
\usepackage{graphicx}
\usepackage{epstopdf}
\usepackage{color}

\begin{document}

\title{Gold Matrix Ghost Imaging}
\author{Xiwei Zhao$^{1,4}$, Xue Wang$^{1,4}$, Wenying Zhang$^{3,4}$, Shanshan Jia$^{1,4}$, Cheng Zhou$^{2,3,4}$, and Lijun Song$^{3,4}$}

\address{$^1$ School of Science, Changchun University, Changchun 130022, People¡¯s Republic of China\\
$^2$ Center for Quantum Sciences and School of Physics, Northeast Normal University, Changchun 130024,People¡¯s Republic of China\\
$^3$ Institute for Interdisciplinary Quantum Information Technology, Jilin Engineering Normal University, Changchun 130052, People¡¯s Republic of China\\
$^4$ Jilin Engineering Laboratory for Quantum Information Technology, Changchun 130052, People¡¯s Republic of China\\ }
\ead{zhoucheng91210@163.com and ccdxslj@126.com}

\vspace{10pt}
\begin{indented}
\item[]November 2019
\end{indented}

\begin{abstract}
\textbf{Light field modulation matrix is closely related to the quality of reconstructed image in ghost imaging. The orthogonality light field modulation matrix with better noise immunity and high quality reconstructed image is urgently needed in the practical application of ghost imaging. In this work, we propose a Gold matrix ghost imaging method with excellent imaging and anti-noise performance, which is proud of the pseudo-randomness and orthogonality of the deterministic Gold matrix. Numerical simulation and experimental results show that the Gold matrix has the characteristics of both random speckle and Hadamard matrix, and the reconstructed images are superior to the two modulation matrixes under the condition of noise, which shows better imaging robustness. }
\end{abstract}

\vspace{2pc}
\submitto{\JOPT}
\maketitle
\ioptwocol
\section{Introduction}

Ghost imaging (GI) is an indirect imaging technique which reconstructs an entire image by correlation of intensity fluctuations. Compared to conventional optical imaging, GI has advantages in many applications, such as gas imaging\cite{gas}, infrared imaging \cite{infrared}, multispectral imaging \cite{multispectral1,multispectral2,multispectral3}, microscopy \cite{microscopy}, 3D imaging \cite{3D1,3D2,3D3,3D4,3D5,3D6}, imaging through scattering media \cite{scattering1,scattering2}, X-ray diffraction tomography \cite{X1,X2}, etc.

For realistic application, fast and high quality GI is an important and urgent task. It is generally known that the imaging speed and imaging quality using different red spatial light modulation patterns vary widely, which determines whether high-quality GI in real-time can be achieved. According to the requirement of scientific research and practical application, commonly used patterns mainly include random speckle patterns (RSP) and deterministic orthogonal basis patterns (DOP). Among them, RSP in conventional dual-path pseudothermal light GI is the pattern of negative exponential distribution points generated by laser beam passing through the rotating ground glass \cite{thermal0,thermal1,thermal2,random}, which is widely used in remote sensing due to the better robustness and high damage threshold of the generated device \cite{damagethreshold}. As a matter of fact, due to the inevitable correlation of RSP, a large number of measurements (far exceeding the number of pixels) are required to achieve better imaging quality \cite{Wu}. In addition to further optimizing RSP \cite{orderofspeckle} or using imaging algorithm \cite{PGI,SMGI}, GI via RSP will be difficult to meet practical application.

Fortunately, due to the orthogonal completeness of most DOP (eg. Fourier basis pattern and Hadamard basis pattern), GI based on DOP can perfectly reconstruct the target image at full sampling (equal to pixels), which greatly reduce the number of measurements. The proposal and experimental implementation of computational ghost imaging (CGI) \cite{2008s} with spatial light modulation equipment (eg. phase-only liquid crystal spatial light modulator (PL-SLM) \cite{2008s}, digital light projector (DLP) \cite{multispectral1}, digital micromirror device (DMD) \cite{DMD}, etc.) makes CGI based on DOP possible. Here, DMD based on micro electro-mechanical systems technology is often used as the spatial light modulation equipment of CGI due to its frame rates far higher than PL-SLM and DLP\cite{DMDbetter}. However, DMD is a pure binary amplitude modulator, which makes it impossible to accurately realize the light field modulation of the grayscale mode Fourier basis pattern \cite {FSI1, FSI2, FSI3,errors}. That means the CGI based on Fourier basis pattern still needs a long imaging time.

Obviously, binary DOP is the key to improve imaging speed in CGI based on DMD, which happens to be satisfied by Hadamard basics pattern. Therefore, to further promote the practical application, the optimization coding of Hadamard basics pattern for CGI was considered \cite{Sun,Zhou,Li,Yu,Wu}. Sun \textit{et al} proposed a Russian Dolls ordering of the Hadamard basis, which uses less computational resources to achieve the imaging quality of conventional methods\cite{Sun}. Then, a multi-resolution CGI method using optimized Hadamard basis was reported by Zhou \textit{et al} \cite{Zhou}. Subsequently, Yu \textit{et al} proposed origami and cake-cutting Hadamard basis sort to realize sub-Nyquist imaging \cite{Yu}. Similarity, Li \textit{et al} utilized Hadamard basis optimized reordering method based on Haar wavelet transform to realize high speed imaging\cite{Li}. Noteworthy, CGI based on Hadamard basis is sensitive to noise, but the method to improve the imaging quality under noise condition has not been discussed. Wang \textit{et al} \cite{Wang} proposed differential imaging of detection signal by using Hadamard basis pairs ($+1$ and $-1$) to improve the robustness against noise. But unavoidably, in the active CGI experimental system, it needs to consume twice the number of measurements.

Hence, DOP with better noise resistance and high imaging efficiency are especially needed in GI applications. Here, we demonstrate a new DOP ghost imaging method for high-speed and high-quality imaging. Because the method is based on Gold matrix, which the deterministic Gold matrix has both orthogonality and pseudo-randomness, the proposed method has some advantages: (1) Due to the orthogonality of the Gold matrix, perfect reconstructed image can be obtained with full sampling; (2) Because of the pseudo-randomness of the Gold matrix, the reconstructed image can be better than Hadamard basis in noisy environment, which shows better imaging robustness; (3) The Gold matrix is a binary matrix with a distribution of 0 and 1, and no measurement matrix pairs and differential signal processing is required in imaging. Therefore, Gold matrix GI based on ultra-high-speed DMD can satisfy both active and passive fast imaging.

\section{Theoretical model}

In this section, we introduce the generating rules of Gold matrix for spatial light field modulation in GI. The randomness and orthogonality of Gold matrix are analyzed by the generating theory of Gold matrix and characteristic matrix, and the comparison with random speckle and Hadamard matrix is further made.

\subsection{Gold matrix}
The Gold sequence and m-sequence are the most wildly used codes in spread spectrum communication systems \cite{LFSR}. Recently, Li \textit{et al}\cite{Lihao} realized high resolution and large range two-dimensional measurements by using two-dimensional Gold matrix composed of Gold sequences. And the Gold matrix has some pseudo-random properties and can be designed. For this reason, we use m-sequences to generate a more general Gold sequence and use it to optimize and design a new Gold matrix with pseudo-randomness and orthogonality for GI. The generating rules of the new Gold matrix can be summarized as two steps:

\textit{First, generating binary m-sequence.} The m-sequence has several important randomness properties and are known as pseudo-random sequence. Next, the generation method of m-sequence is introduced, which is obtained by using linear feedback shift registers (LFSR) \cite{LFSR}. The length of the LFSR is equal to the order of the m-sequence and the generation rule is summarized as the corresponding polynomial, which is primitive polynomial and can be expressed as:

\begin{equation}
f(x) = c_0 + c_1x + c_2x^2 + \cdots + c_kx^k.
\end{equation}

The coefficient of primitive polynomial is always arranged from high to low as:

\begin{equation}
({c_k},{c_{k - 1}}, \cdots,c_0),~{c_i} \in \left\{ {0,1} \right\},~i = 0,1, \cdots ,k.
\end{equation}

 A universal state of LFSR is:

\begin{equation}
({a_0},{a_1}, \cdots, {a_{k - 1}}),~{a_i} \in \left\{ {0,1} \right\},~i = 0,1, \cdots k - 1.
\end{equation}

There is a clock in the LFSR to indicate the number of changes in the state. The number of clock turn is expressed in $l,~l=0,1,2,\cdots,2^{k}$. The clock does not turn when $l=0$, i.e., the initial value of the LFSR can be expressed as:
\begin{equation}\label{EQ1}
{({a_0},{a_1}, \cdots, {a_{k - 1}})^{(0)} = \{{L(i)|i=0,1,\cdots,k-1}\}},
\end{equation}
where $L\left( i \right)$ represents the value (0 or 1) of the $ith$ position in the LFSR. Turn the clock once ($l=1$), the output value of LFSR is $L(0)=(a_0)^{(1)}$.

Hence, the residual part value of the LFSR shifts and a new LFSR is obtained:

\begin{equation}
{\left( {{a_0},{a_1}, \cdots, {a_{k - 2}}} \right)^{(1)}} = {({a_1},{a_2}, \cdots, {a_{k - 1}})^{(0)}},
\end{equation}
and the highest value $(a_{k - 1})^{(1)}$ will be determined by the  coefficient of primitive polynomial and the previous state:
\begin{equation}
{\left( {a_{k - 1}} \right)^{(1)}} = {c_1} \cdot {\left( {{a_{k - 1}}} \right)^{(0)}} \oplus {c_2} \cdot {\left( {{a_{k - 2}}} \right)^{(0)}} \oplus  \cdots  \oplus {c_k} \cdot {\left( {{a_0}} \right)^{(0)}}.
\end{equation}

After the clock $2^k-1$ turns, the LFSR outputs $2^k-1$ values $[(a_0)^{(1)},(a_0)^{(2)},\cdots,(a_0)^{(2^k-1)}]$ in total. These output values form a m-sequence $\{{X(i)|i=0,1,\cdots,2^{k}-2}\}$.

\textit{Second, generating binary general Gold matrix.} In general, to generate a Gold sequence, m-sequence preferred pair needs to be obtained first. However, we found that only two m-sequences generated by shift registers defined by different primitive polynomials can achieve perfect reconstruction in full sampling in GI. That's why we call it 'general' Gold sequence to distinguish it.

The following describes the method of generating a $k-order$ Gold matrix using m-sequences. Before starting, we need to get two different sets of $k-order$ m-sequences $\{{X(i)|i=0,1,\cdots,2^{k}-2}\}$ and $\{Y(j)|{j = 0,1,\cdots,{2^k}-2}\}$.

Then, the m-sequence $\{{X(i)|i=0,1,\cdots,2^{k}-2}\}$ is copied as a row to get the matrix A, i.e., each row of matrix $A$ is identical to the m-sequences $X$ and can be expressed as:
\begin{equation}\label{amatrix}
 A = \left[ {\begin{array}{*{20}{c}}
{X(0)}&{X(1)}& \cdots &{X({2^k} - 2)}\\
{X(0)}&{X(1)}& \cdots &{X({2^k} - 2)}\\
 \vdots & \vdots & \ddots & \vdots \\
{X(0)}&{X(1)}& \cdots &{X({2^k} - 2)}
\end{array}} \right].
\end{equation}

Put slightly differently, the m-sequence\\ $\{{Y(j)|j=0,1,\cdots,2^{k}-2}\}$ forms the matrix B by cyclic shifting:
\begin{equation}\label{bmatrix}
 B = \left[ {\begin{array}{*{20}{c}}
{Y(0)}&{Y(1)}& \cdots &{Y({2^k} - 2)}\\
{Y({2^k} - 2)}&{Y(0)}& \cdots &{Y({2^k} - 3)}\\
 \vdots & \vdots & \ddots & \vdots \\
{Y(1)}&{Y(2)}& \cdots &{Y(0)}
\end{array}} \right].
\end{equation}
Subsequently, a new matrix C is obtained by XOR operation of matrix A and matrix B. According to Eq.~\ref{amatrix} and Eq.~\ref{bmatrix}, the matrix C can be expressed as follows in Eq.~\ref{cmatrix}. If the m-sequence preferred pair is used in generating the matrix $A$ and the matrix $B$, then each row of the matrix $C$ (Eq.~\ref{cmatrix}) is a Gold sequence.

\begin{table*}[t]
\begin{equation}\label{cmatrix}
 C = \left[ {\begin{array}{*{20}{c}}
{X(0) \oplus Y(0)}&{X(1) \oplus Y(1)}& \cdots &{X({2^k} - 2) \oplus Y({2^k} - 2)}\\
{X(0) \oplus Y({2^k} - 2)}&{X(1) \oplus Y(0)}& \cdots &{X({2^k} - 2) \oplus Y({2^k} - 3)}\\
 \vdots & \vdots & \ddots & \vdots \\
{X(0) \oplus Y(1)}&{X(1) \oplus Y(2)}& \cdots &{X({2^k} - 2) \oplus Y(0)}
\end{array}} \right].
\end{equation}
\end{table*}

Based on the matrix $C$, we have further optimized the design to get the Gold matrix, which consists of the matrix $C$, the m-sequence $X$ and a column vector of length $2^k$ with all values being $1$, as shown in Eq.~\ref{goldmatrix}. In order to realize the spatial light field modulation of GI, each row of the Gold matrix (Eq.~\ref{goldmatrix}) is converted from a row vector of length $2^{k/2}\times 2^{k/2}$ , which is obtained by reshaping the $sth$ light field ${I^{(s)}}(x,y)$.
\begin{table*}[t]
\begin{equation}\label{goldmatrix}
 Gold = {\left[ {\begin{array}{*{20}{c}}
1&{X(0) \oplus Y(0)}&{X(1) \oplus Y(1)}& \cdots &{X({2^k} - 2) \oplus Y({2^k} - 2)}\\
1&{X(0) \oplus Y({2^k} - 2)}&{X(1) \oplus Y(0)}& \cdots &{X({2^k} - 2) \oplus Y({2^k} - 3)}\\
1& \vdots & \vdots & \ddots & \vdots \\
 \vdots &{X(0) \oplus Y(1)}&{X(1) \oplus Y(2)}& \cdots &{X({2^k} - 2) \oplus Y(0)}\\
1&{X(0)}&{X(1)}& \cdots &{X({2^k} - 2)}
\end{array}} \right]}.
\end{equation}
\end{table*}

%In order to be loaded onto the DMD, the gold matrix needs to be further processed. The size of the k-order gold matrix is ${2^k} \times {2^k}$. In gold matrix GI, the reshaped rows of gold matrix are used to generate gold matrix pattern ${I^s}(x,y)$, expressed as
%\begin{equation}
%{I^s}(x,y) = reshape(Gold(s)),s = 1,2 \cdots {2^k}
%\end{equation}
%Where $Gold(s)$ represents the $ith$ row of gold matrix and $(x,y)$ represents the pixel coordinates. The function $reshape()$ represents converting a one-dimensional vector into a matrix. At this point, the completion of the generation of a series of ${I^{(s)}}(x,y)$ and only $0$ and $1$ matrix can be directly loaded into the DMD.

\subsection{GI image reconstruction and the normalized characteristic matrix}

In the framework of GI, the reflection or transmission coefficient of target object $O(x,y)$ can be obtained by computing the correlation between ${I^{(s)}}(x,y)$ and $D^{s}$ :

\begin{eqnarray}\label{eqtgi}
%\begin{aligned}
O_{GI}(x,y) &= \frac{1}{K}\sum^{K}_{s=1}[(D^{(s)}-\langle D^{(s)}\rangle)\cdot\nonumber\\
&(I^{(s)}(x,y)-\langle I^{(s)}(x,y)\rangle)],
%\end{aligned}
\end{eqnarray}
where $\left\langle {{D^{(s)}}} \right\rangle  = \frac{1}{K}\sum_{s=1}^K{D^{(s)}}$ and $\left\langle {I^{(s)}(x,y)} \right\rangle  = \frac{1}{K}\sum_{s=1}^K{I^{(s)}(x,y)} $. Among them, the $s$th bucket signal $D^{(s)}$ can be further expanded as follows:

\begin{equation}\label{DI}
 {D^{(s)}} = \int\int{{{I^{(s)}}(x,y)O(x,y)}}\mathrm{d}x\mathrm{d}y,
\end{equation}
where, $s=1,2,3,\cdots,K$.

We can transform ${I^{(s)}}(x,y)$ (dimensions $m\times n$ ) of $K$ measurements into a $K\times N (N=m\times n)$ matrix:
\begin{eqnarray}
 M = {\left[ {\begin{array}{*{20}{c}}
{I_{}^{(1)}(1,1)}&{I_{}^{(1)}(2,1)}& \ldots &{I_{}^{(1)}(m,n)}\\
{I_{}^{(2)}(1,1)}&{I_{}^{(2)}(2,1)}& \ldots &{I_{}^{(2)}(m,n)}\\
 \vdots & \vdots & \ddots  & \vdots \\
{I_{}^{(K)}(1,1)}&{I_{}^{(K)}(2,1)}& \ldots &{I_{}^{(K)}(m,n)}
\end{array}} \right]}.
\end{eqnarray}
Here, the rows of the matrix $M$ is obtained from the reshaped patterns ${I^{(s)}}(x,y)$, which is to represent all the modes used for sampling a total of $K$ times in the form of a two-dimensional matrix. Thus, Eq.~\ref{eqtgi} can be rewritten into a matrix form:
\begin{eqnarray}\label{mgi}
O_{GI}&=&\frac{1}{K}{(M - I\left\langle {{M}} \right\rangle )^{\rm{T}}}(D - I\left\langle D \right\rangle )\nonumber\\
      &=&\frac{1}{K}{(M - I\left\langle M \right\rangle )^{\rm{T}}}(M - I\left\langle {{M}} \right\rangle )O\\
      &=&\frac{1}{K}{M_{C}}O\nonumber,
\end{eqnarray}
where $D = {\left[D^{(1)},D^{(2)}, \cdots, D^{(K)} \right]^{\rm{T}}}$ denotes a $K\times 1$ column vector and object $O(x,y)$ is represented as $N\times 1$ column vector. ${M_C} = {(M - I\left\langle M \right\rangle )^{\rm{T}}}(M - I\left\langle {{M}} \right\rangle )$ is called the characteristic matrix \cite{wcl}.

The normalized characteristic matrix $M_C^N$, denoted by normalizing the matrix ${M_{C}}$  with its two-norm value of the row where its maximum exists, can be used to analyze the error caused by the pattern itself when reconstructing the image. If the $sth$ row in $M_C^N$ is understood as a weight coefficient, the $sth$ value in the ${O_{GI}}$ can be regarded as a weighted sum of all values in the object $O$. The $sth$ value in the ${O_{GI}}$ is only corresponding to the $sth$ value in the $O$ and only the value of the weight coefficient is the signal weight. For an element in ${O_{GI}}$, ideally it should be equal to the corresponding position in $O$, so if the weight coefficient of other positions (off-diagonal region of $M_C^N$) is not zero, it will bring errors to the reconstruction result. We try to make the characteristic matrix a unit matrix to achieve perfect GI.
\begin{figure*}[htbp]
\centering \includegraphics[width=18 cm]{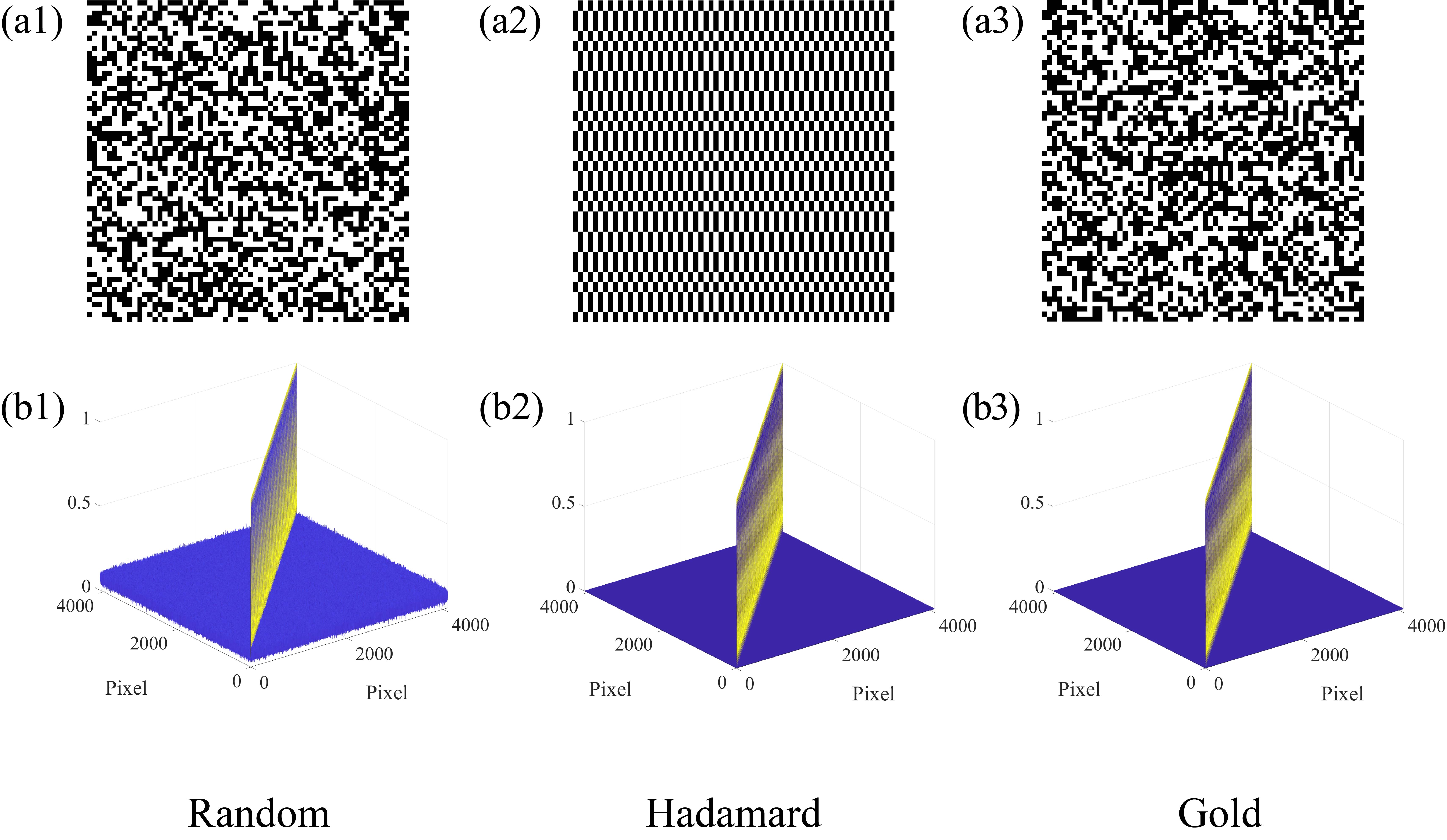}
  \caption{Light field modulation patterns and normalized characteristic matrices. The diagram of the matrix $I(x,y)$ for three kinds of light field modulation patterns: (a1) Random, (a2) Hadamard, (a3) Gold;  The diagram of the matrix $M_C^N$ for three kinds of normalized characteristic matrices: (b1) Random, (b2) Hadamard, (b3) Gold.}
\label{analysis}
\end{figure*}
\par

 In order to analyze and compare the performance of Gold matrix in GI, the typical patterns and the $M_C^N$s of random speckle, Hadamard and Gold matrix are given in Fig.~\ref{analysis}. As a deterministic matrix, there is still pseudo-randomness in each row of the Gold matrix generated using the m-sequences. As shown in Fig.~\ref{analysis}, the pattern of Gold matrix (Fig.~\ref{analysis}(a3)) has a similar structure to the random speckle pattern (Fig.~\ref{analysis}(a1)) and the pseudo-randomness in the row of the Gold matrix still exists after being reshaped. The calculation result of ${M_C^N}$ is shown in Figs.~\ref{analysis}(b1)-\ref{analysis}(b3). Both Hadamard (Fig.~\ref{analysis}(b2)) and Gold matrix (Fig.~\ref{analysis}(b3)) have zero diagonal regions of $M_C^N$, which reflects the orthogonality of the Gold matrix. Thus, it is proved that Gold matrix has both pseudo-randomness and orthogonality. In addition, if the values in the Gold matrix are mapped from (1,0) to (1,-1), the Gold matrix is strictly orthogonal.

\subsection{Performance evaluation}
In order to objectively evaluate the imaging quality of Gold matrix GI, mean square error (MSE) and peak signal-to-noise ratio (PSNR) are used, which are defined as:
\begin{equation}
MSE = \frac{1}{{m \times n}}\sum\limits_{x = 1}^m {\sum\limits_{y = 1}^n {{{\left[ {O - {O_{{GI}}}} \right]}^2}} },
\end{equation}
and
\begin{equation}
PSNR = 10 \times {\log _{10}}\left[ {\frac{{\max Va{l^2}}}{{MSE}}} \right],
\end{equation}
where $max Val$ is the maximum possible pixel value of the image.
\section{Simulation and experimental results}
In order to test the feasibility of the Gold matrix GI, we performed simulations and experiments. The effects of noise and three kinds of light field on imaging quality are discussed.
\subsection{Simulation results}
\begin{figure}[tbp]
\centering \includegraphics[width=8 cm]{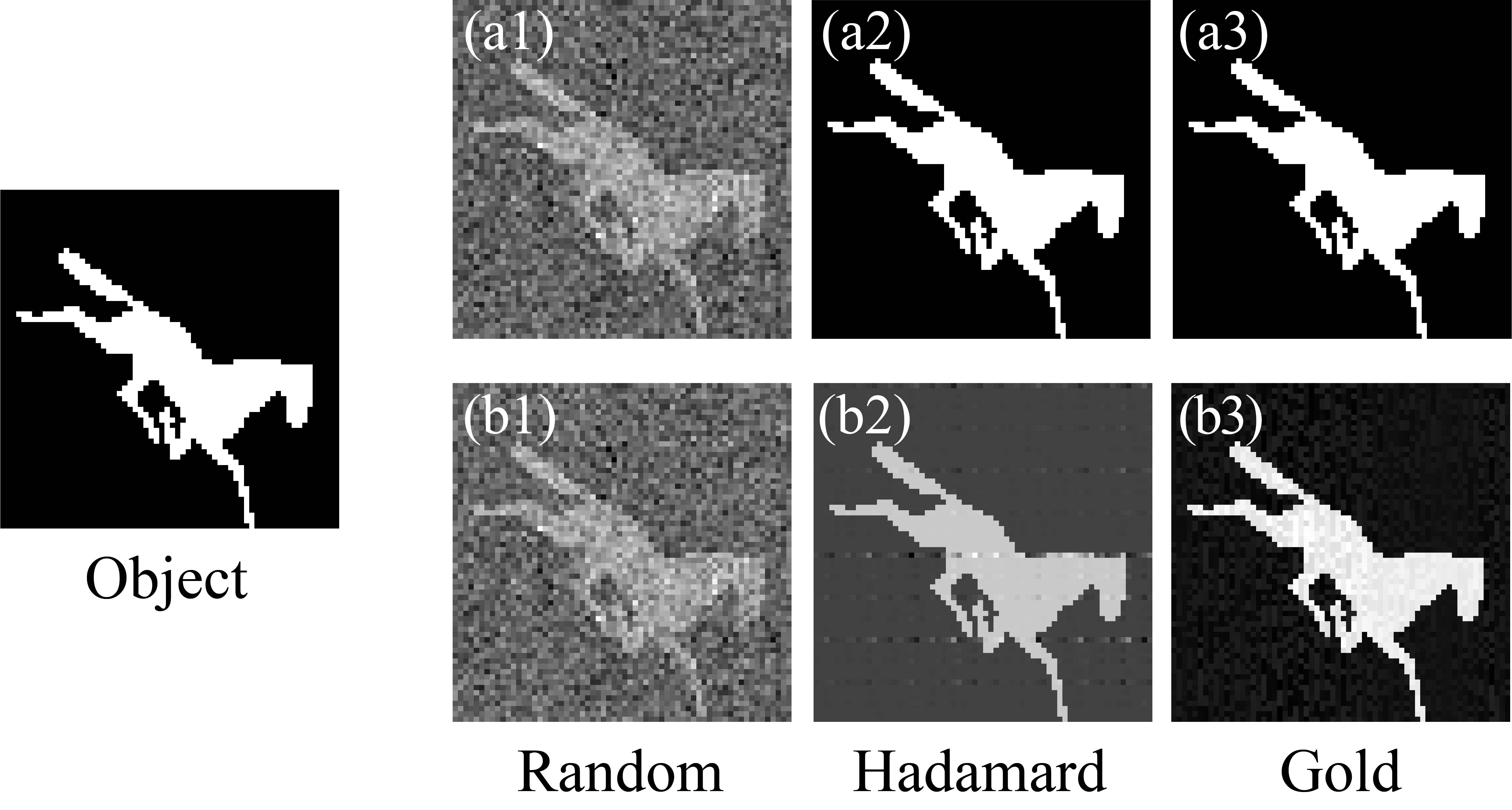}
  \caption{Numerical experimental demonstration of a binary horse for three kinds of patterns when $K=4096$. Ideal environment: (a1) Random, (a2) Hadamard, (a3) Gold; Noisy environment: (b1) Random, (b2) Hadamard, (b3) Gold.}
\label{fig2}
\end{figure}

 We performed two sets of simulations in ideal environment and noisy environment. The results are shown in Fig.~\ref{fig2}, where the pixel size of the object is $64\times 64$ and the number of measurements is $4096$. The binary image of the horse and the grayscale image of the house are selected as the simulated object. First, the simulation results of the horse are introduced.
\par

The reconstructed images in ideal environment are shown in Figs.~\ref{fig2}(a1)-\ref{fig2}(a3). The reconstructed images of Hadamard (Fig.~\ref{fig2}(a2)) and Gold matrix (Fig.~\ref{fig2}(a3)) are completely faithful to the object, achieving perfect reconstructions of full sampling. The reconstructed image of random speckle (Fig.~\ref{fig2}(a1)) is blurred. In ideal environment, the reconstructed image of Gold matrix is the same as that of Hadamard and is significantly better than that of random speckle. The reconstructed images are consistent with those predicted by the normalized characteristic matrix.

In order to make the simulation results more accurately reflected in the reconstruction results of the actual imaging environment, considering the ambient lighting, the instability of the light source and the dark current of the detector itself can affect the imaging results, we add some noise into the numerical simulation and the imaging results are shown in Figs.~\ref{fig2}(b1)-\ref{fig2}(b3). The imaging results in a noisy environment show that there is a region of noise concentration in the reconstructed image of Hadamard, and the image information of these regions is seriously damaged, as shown in Fig.~\ref{fig2}(b2). Under the same conditions, the reconstructed image of the Gold matrix can still accurately reflect the object information (see Fig.~\ref{fig2}(b3)). The results show that the Gold matrix has better imaging robustness.

\begin{figure}[tbp]
\centering \includegraphics[width=7 cm]{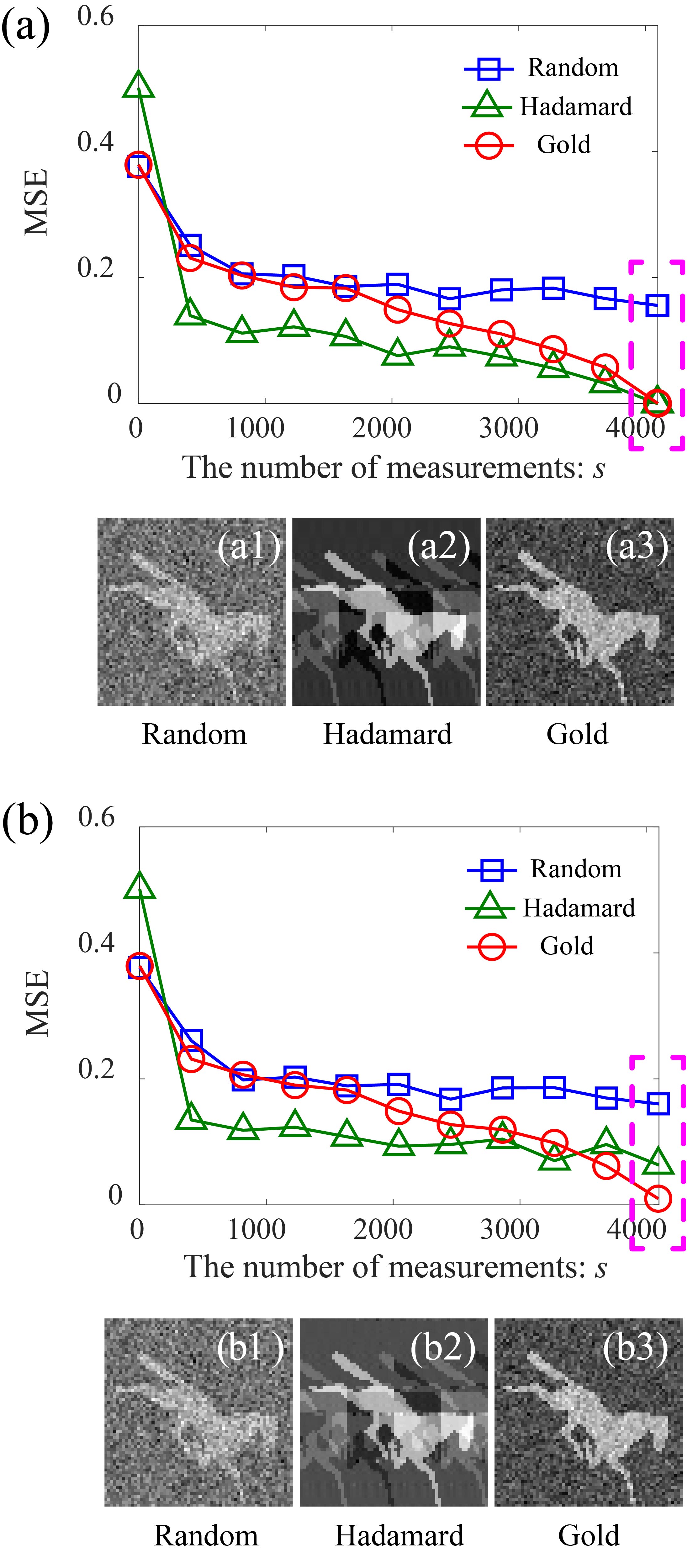}
  \caption{The relationship between MSE and the number of measurements for three kinds of patterns in ideal and noisy environment. Ideal environment: (a) The curve of MSE-$s$; The reconstructed results of (a1) Random, (a2) Hadamard and (a3) Gold at $s=3000$. Noisy environment: (b) The curve of MSE-s; The reconstructed results of (b1) Random, (b2) Hadamard and (b3) Gold at $s=3000$.}
\label{fig33}
\end{figure}
\par

To quantify the image quality, we calculated the MSE of the reconstructed images of the binary image. Here, MSE is inversely proportional to the imaging quality. The MSE calculation results of the reconstructed image under noiseless conditions are shown in Fig.~\ref{fig33}(a). As the number of measurements increases, the MSE of Hadamard and Gold matrix decreases significantly and reaches zero at full sampling and their imaging quality is far superior to that of random speckle. The MSE of the reconstructed images in a noisy environment are shown in Fig.~\ref{fig33}(b). The imaging quality of Hadamard and Gold matrix full sampling is quite different. At the time of full sampling, the MSE of Hadamard rises remarkably, while the MSE of the Gold matrix rises less. This indicates that noise has little effect on the image quality of Gold matrix. The calculation results of MSE show that the Gold matrix has obvious advantages compared with Hadamard in terms of noise immunity. Since there is inevitable noise in the actual imaging environment, superior noise immunity is of great significance for the practical application of the technology.

In addition, at a certain number of measurements (under undersampling), MSE indicates that the imaging quality of Hadamard is better than the other patterns. Hence, we make a concrete comparison of the reconstructed image results. The reconstructed image results of 3000 measurements are shown in Figs.~\ref{fig33}(a1)-\ref{fig33}(a3) and Figs.~\ref{fig33}(b1)-\ref{fig33}(b3). The reconstructed image of Hadamard is overlapped, thereby destroying the image. The reconstructed image of the random speckle and the Gold matrix are normal. Therefore, Gold matrix also has better imaging performance under undersampling condition.

 \begin{figure}[tbp]
\centering \includegraphics[width=8 cm]{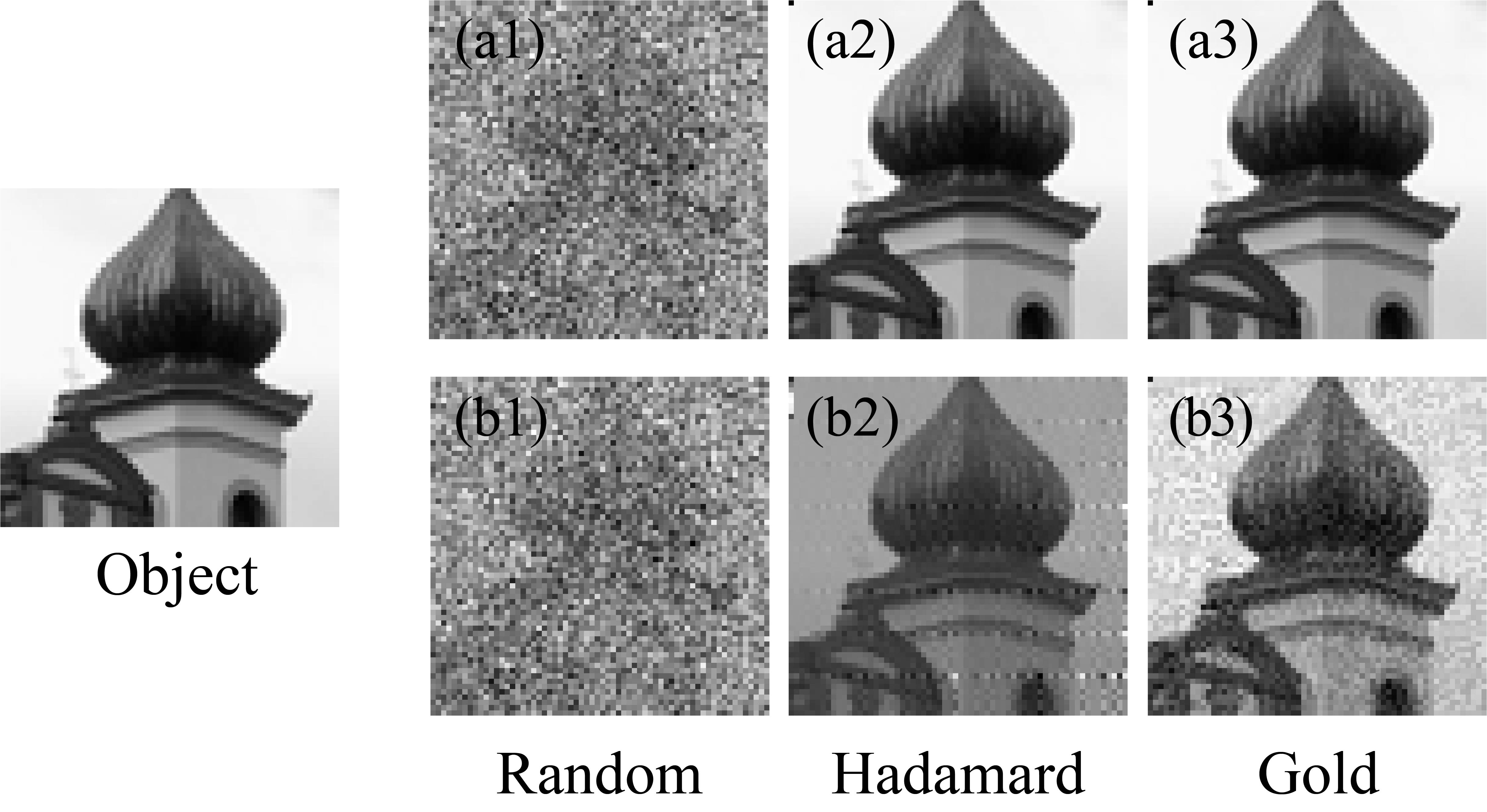}
  \caption{Numerical experimental demonstration of a grayscale house for three kinds of patterns when $K=4096$. Ideal environment: (a1) Random, (a2) Hadamard, (a3) Gold; Noisy environment: (b1) Random, (b2) Hadamard, (b3) Gold.}
\label{fig44}
\end{figure}
\par

 \begin{figure}[tbp]
\centering \includegraphics[width=7 cm]{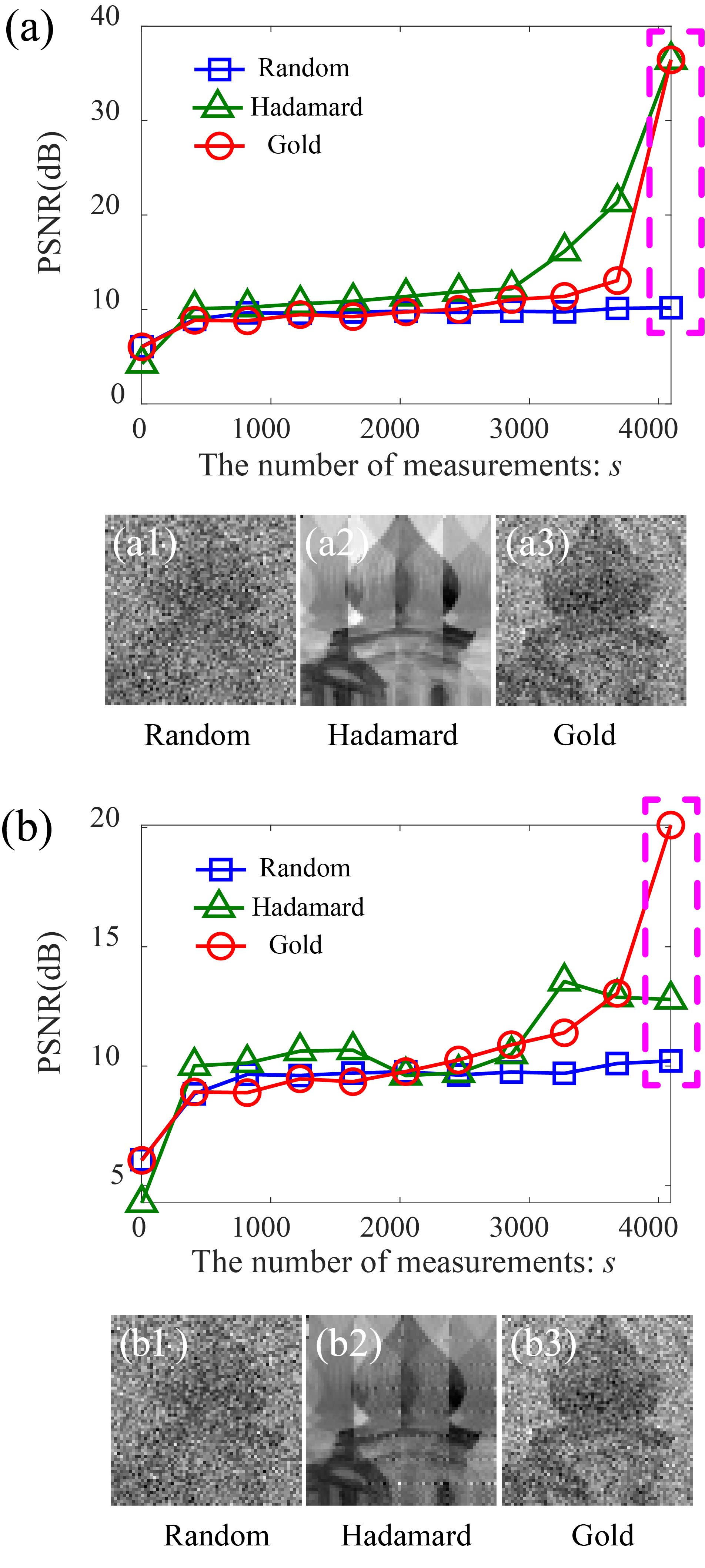}
  \caption{The relationship between PSNR and the number of measurements for three kinds of patterns in ideal and noisy environment. Ideal environment: (a) The curve of PSNR-$s$; The reconstructed results of (a1) Random, (a2) Hadamard and (a3) Gold at $s=3000$. Noisy environment: (b) The curve of PSNR-s; The reconstructed results of (b1) Random, (b2) Hadamard and (b3) Gold at $s=3000$.}
\label{fig55}
\end{figure}
\par

The simulation results of the house are shown in Fig.~\ref{fig44} and Fig.~\ref{fig55}, which is consistent with the simulation results of the horse. The simulation results preliminarily proved the feasibility of Gold matrix GI. In addition, the simulation results in a noisy environment show that in the actual environment, the reconstructed image quality using the Gold matrix pattern may be better than the commonly used random speckle pattern and Hadamard pattern.

\subsection{Experimental results}

In order to verify the simulation results and further prove the feasibility of Gold matrix GI, we performed imaging experiments.

 \begin{figure}[tbp]
\centering \includegraphics[width=8 cm]{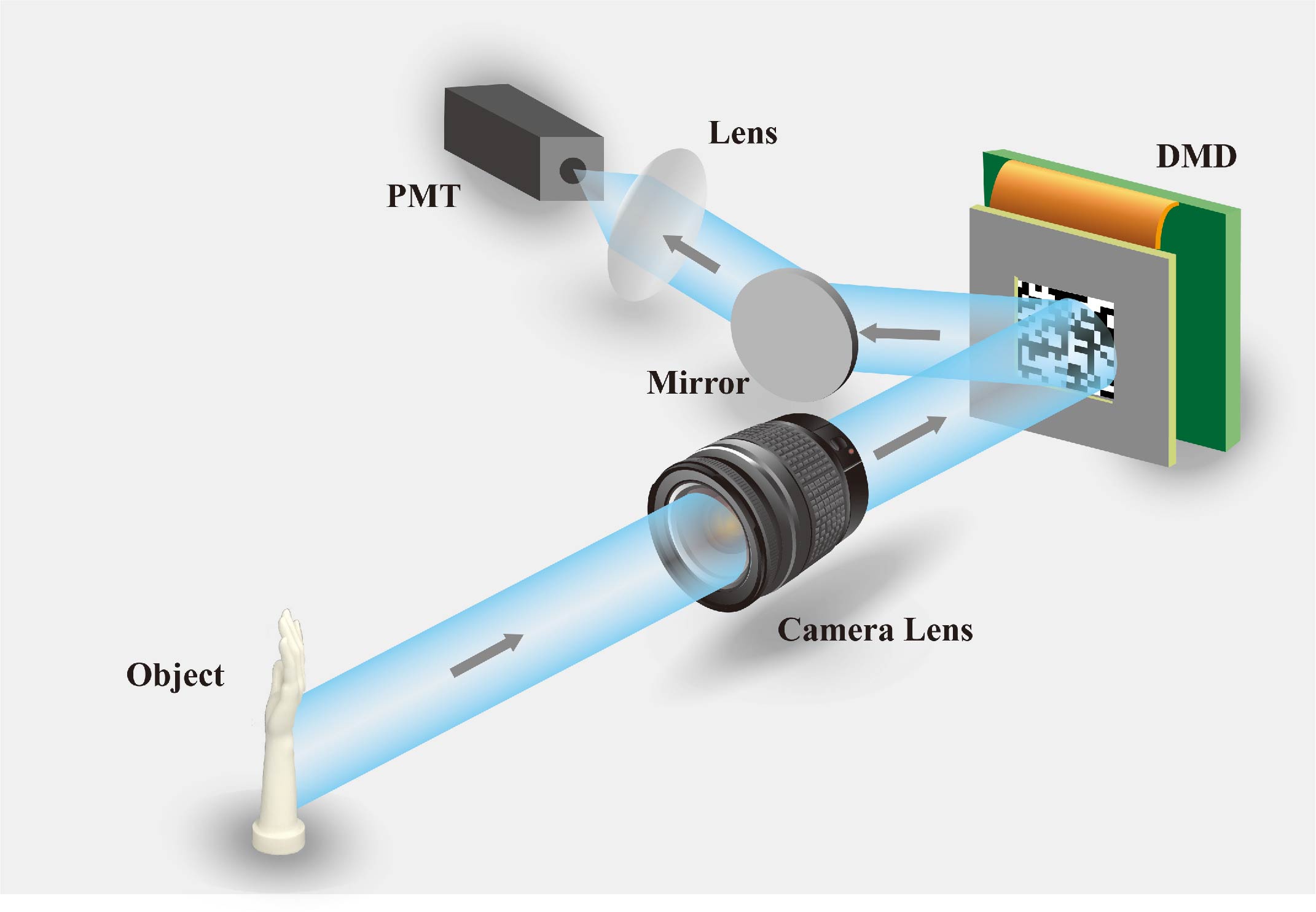}
  \caption{The experiment system diagram of computational ghost imaging.}
\label{fig66}
\end{figure}
\par

Fig.~\ref{fig66} illustrates the structure of the experimental system. The target object is a model of an all-white hand with a size of $35\textrm{cm}\times 10
\textrm{cm}$ , which is located ~$5$ meters from the imaging system. The imaging system consists of a camera lens, a DMD, a total reflection mirrors, a collection lens and a photomultiplier tube. The light (cold white LED)  illuminates the object. Then, the reflected light is illuminated onto the DMD through the camera lens. The applied DMD consists of $1024\times 768$ micromirrors, each of which can be switched between two directions of  $ \pm {12^ \circ }$, corresponding to $1$ and $0$. It displays a pre-loaded sequence of binary patterns $(64 \textrm{pixel}\times 64 \textrm{pixel})$ at a rate of 1K pattern/second which are used to modulate the light field. A current output PMT (Hamamatsu H10721-01)  is placed on the reflection direction of DMD to make the measurement of total echo signal.

 \begin{figure}[tbp]
\centering \includegraphics[width=8 cm]{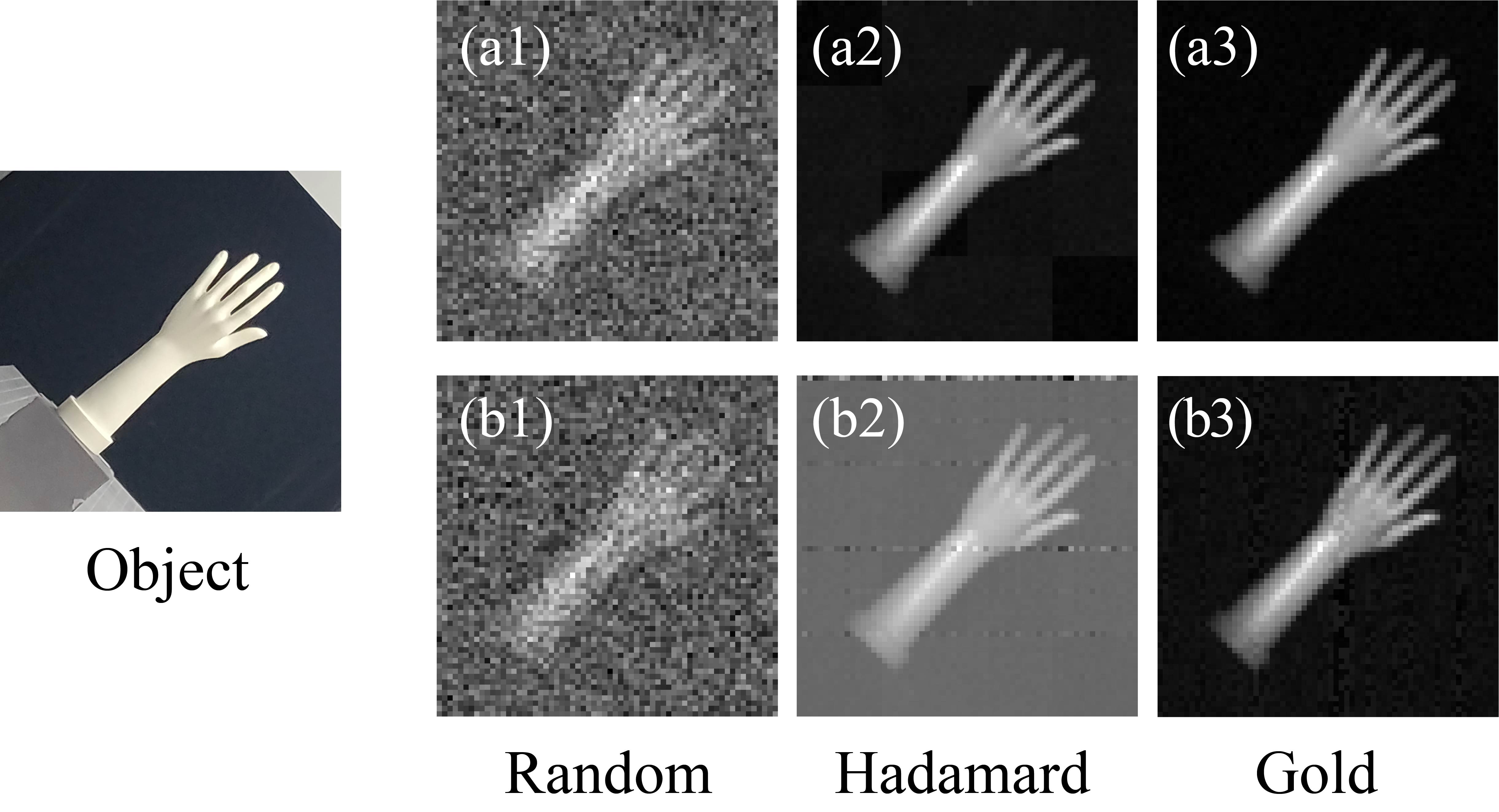}
  \caption{Experimental demonstration of a hand model for three kinds of patterns when $K=4096$. Ideal environment: (a1) Random, (a2) Hadamard, (a3) Gold; Noisy environment: (b1) Random, (b2) Hadamard, (b3) Gold.}
\label{fig77}
\end{figure}
\par

We performed imaging experiments in ideal environment and noisy environment. The experimental results are shown in Fig.7, which is basically consistent with the simulation results. As shown in Figs.~\ref{fig77}(a1)-\ref{fig77}(a3), in the ideal environment, the image quality of random speckle is poor and the reconstructed images of Hadamard and Gold matrix are basically faithful to the original object. Due to the inevitable noise, Hadamard's reconstructed image showed a small amount of noise, while Gold matrix did not. The experimental results in the noisy environment as shown in Figs.~\ref{fig77}(b1)-\ref{fig77}(b3), the reconstructed images of Hadamard and Gold matrix are still better than random speckle, but in the reconstructed image of Hadamard, there are multiple strips of noise regions, and the edges of the objects in these regions have large errors, and the overall contrast of the image is reduced. The reconstructed image of the Gold matrix is significantly better than that of Hadamard. The edge of the object is clear, the gray value of the object is more accurate. The experimental results prove the feasibility of the Gold matrix GI, and the comparison of the full-sample reconstructed images in the noisy environment shows that the Gold matrix ghost imaging has better noise immunity than the Hadamard.

\section{Conclusion}
We have proposed and demonstrated a new deterministic orthogonal matrix with randomness imaging method, which is called Gold matrix ghost imaging. The numerical and experiments show that the imaging quality using Gold matrix is significantly better than that using random speckle. Compared with Hadamard, Gold matrix can also achieve perfect reconstruction of objects in full sampling and the reconstructed image quality is less affected by noise, and still provides better image quality in the actual environment. In addition, Gold matrix is also easy to be generated using LFSR and match DMD usage. We believe that gold matrix GI performs better than existing solutions and will further promote the practical use of GI.

\section*{Funding}
This work is supported by the Project of the Special Funds for Provincial Industrial Innovation in Jilin Province (Grant No. 2018C040-4, 2019C025).

%\begin{thebibliography}{99}
%\end{thebibliography}

\end{document}